\documentstyle[aps,preprint]{revtex}  
\begin{document}
\draft
\tighten
\preprint{\vbox{ \hfill YUMS 97-10 \\ 
\null \hfill  SNUTP 97-048 }}
\title{Calculation of $g_{\pi{\rm NS}_{11}}$ and 
$g_{\eta{\rm NS}_{11}}$ Couplings in QCD Sum Rules}
\author{Hungchong Kim\footnote{E-mail:hung@phya.yonsei.ac.kr} and
Su Houng Lee\footnote{E-mail:suhoung@phya.yonsei.ac.kr}}
\address{ Department of Physics, Yonsei University, Seoul, 120-749, Korea}
\date{\today}
\maketitle
\begin{abstract}
We calculate the coupling constants, $g_{\pi{\rm NS}_{11}}$ and
$g_{\eta{\rm NS}_{11}}$,
using QCD sum rules in the presence of an external meson field.
A covariant derivative is
introduced within the S$_{11}$ interpolating field so that in the 
nonrelativistic limit the field dominantly reduces to two quarks in the s-wave
state and one quark in the p-wave state.  Our result for the couplings  
obtained by further making use of  the soft-meson theorem qualitatively 
agrees with its phenomenological value extracted from
the S$_{11}$(1535) decay width.  The prediction for the
couplings however depend on the value of quark-gluon condensate,
$\langle {\bar q} g_s \sigma \cdot {\cal G} q \rangle$, which is also 
important in the calculation of the S$_{11}$(1535) mass itself within the 
sum rule approach.  
\end{abstract}  
\vspace{20pt}

The study of the s-channel S$_{11}$(1535) resonance is particularly 
interesting as it is believed to dominate $\eta$-production on a nucleon 
in the electromagnetic or hadronic probes. A number of experiments of 
$\eta$-production are underway or planned at MAMI (Mainz)~\cite{mami}, 
ELSA (Bonn)~\cite{elsa}
and TJNAF.  In relation to this, there have been a number of theoretical works
on the $\eta$-production within effective models and on the properties of 
 S$_{11}$(1535) resonance itself ~\cite{eta_the}.

In a previous work, we have proposed a new interpolating 
field to study the spectral properties of S$_{11}$(1535) based on
the conventional QCD sum rule approach~\cite{hung}.  In that work,
the S$_{11}$ interpolating field contains the covariant derivative
so that, in the nonrelativistic limit, it 
reduces to the quark configuration
as suggested by the nonrelativistic quark model~\cite{isgur} or
the bag model~\cite{chodos}. 
Using this current,
we were able to predict the mass of S$_{11}$ close to its empirical value,
although the barely known quark-gluon 
condensate $\langle {\bar q} g_s \sigma \cdot {\cal G} q \rangle$ turned out
to be the main ingredient responsible for the splitting from the nucleon.
Similar approach has been successfully applied 
to $\Lambda$(1405)~\cite{lambda}.

As a further test of this interpolating field, it is interesting to calculate
the coupling constants, $g_{\pi{\rm NS}_{11}}$ and 
$g_{\eta{\rm NS}_{11}}$, in the QCD sum rule approach. If successful, it will
give us further insights to the relation between the properties of hadrons and
the structure of QCD.
Phenomenologically,
one obtains $g_{\pi{\rm NS}_{11}}\sim 0.7$ and 
$g_{\eta{\rm NS}_{11}}\sim 2$ from fitting the partial decay widths of the
S$_{11}$ to each mode, both of which are $\Gamma \sim 70$ MeV, by assuming
a constant coupling between the three states involved. 

Recently, Oka, Jido and Hosaka~\cite{oka} [OJH] have calculated the coupling 
constants using the S$_{11}$
interpolating field without the covariant derivative. Their point of
view is that the S$_{11}$ current can be constructed from the usual 
nucleon interpolating field by tuning the parameter
$t$ which appears in the linear combination of the two possible currents
for the nucleon~\cite{jido}.  They found that at $t=0.8$, their current 
predicts the S$_{11}$ mass reasonably well. Using this kind of current for 
S$_{11}$, $J_{N^*} (t=0.8)$,
and the usual {\it Ioffe} current for nucleon, $J_N (t=-1)$, they looked 
at the Dirac structure proportional to
${\gamma_5\not\!q}$ of the two-point correlation function between the vacuum
and, for example,  a pion state in the soft-pion limit:
\begin{equation}
\Pi^\pi (q) = i \int d^4x e^{iq \cdot x} \langle 0 | T J_{N^*} 
(x) {\bar J}_N (0) | \pi^0(0) \rangle \ .
\label{corr}
\end{equation}
They considered the ${\gamma_5\not\!q}$ structure because, at this structure,
the nucleon contribution vanishes and the contribution from
S$_{11}$, which contains the coupling,  dominates over the ones from other 
resonances such as the roper.  
However, in this approach, one can use the soft-pion theorem combined with
the transformation properties of
\begin {equation}
[Q_5, J_N] = i \gamma_5 J_N 
\end{equation}
to prove that the correlator equals to
\begin {equation}
\Pi^\pi (q) \sim \int d^4x e^{i q\cdot x} \{\gamma_5, \langle 0 | T J_{N^*} 
(x) {\bar J}_N (0) | 0 \rangle\}\ .
\label{c_soft}
\end{equation}
This indicates that the structure proportional to ${\gamma_5\not\!q}$ is 
identically zero, because the vacuum expectation value of the two-point function
$\langle 0 | T J_{N^*} (x) {\bar J}_N (0) | 0 \rangle \sim {A\not\!q + B}$.
Therefore, OJH concluded that $g_{\pi {\rm NS}_{11}}$
is zero. Note however that this is just a consequence of the nucleon current
having no preferred direction.
To estimate the nonzero value of $g_{\pi {\rm NS}_{11}}$,
one could consider  
the structure proportional to $\gamma_5$ in this
limit and, in principle, one could construct the sum rule to calculate
$g_{\pi {\rm NS}_{11}}$ coupling.   However, in this structure, the nucleon
double pole term with $g_{\pi {\rm NN}}$ appears in the phenomenological side
and there is no well-defined
way of separating out the contribution only from $g_{\pi {\rm NS}_{11}}$ unless
one enters the value of $g_{\pi {\rm NN}}$ as an input. Also the contribution
from other resonances could be large in contrast to the case for
the ${\gamma_5\not\!q}$ structure.

One of the difficulties above is that the usual nucleon interpolating field is
not a parity eigenstate so that even the current  $J_{N^*} (t=0.8)$ couples 
not only to
the negative parity states but also to the positive parity nucleon state.  
A way out could be to construct a new interpolating field that couples 
strongly to S$_{11}$ but does not
couple to the nucleon.  Such a current was constructed by us before\cite{hung}.
Here, one introduces the covariant derivative within the current,
\begin{eqnarray}
J_{N^*} (t)=2 \epsilon_{abc}[\ t\ (u_a^{\rm T} C (z \cdot D)d_b)
\gamma_5 u_c
+(u_a^{\rm T} C \gamma_5 (z \cdot D) d_b) u_c]\ ,
\label{nstar}
\end{eqnarray}
where $z^\mu$ is the auxiliary space-like vector which is  orthogonal
to the four momentum carried by S$_{11}$.  Therefore, in the rest frame of
S$_{11}$, $z \cdot D$ reduces to the derivative in the space direction.
Due to this specific structure, this interpolating field couples to
S$_{11}$ via
\begin{equation}
\langle 0 | J_{N^*}(0) | {\rm S}_{11} (q,s) \rangle =
-i \gamma_5 \not \!z \lambda_{N^*} (t) \ u_{N^*}(q,s)\ .
\label{mat}
\end{equation}
Here $u_{N^*}(q,s)$ represents the S$_{11}$ spinor with momentum $q$ and 
spin $s$.   In principle, $J_{N^*}$ also couples to nucleon.  
However, the coupling strength to the nucleon can be tuned to zero using
the finite energy sum rule~\cite{hung}, which is 
obtained from a sum rule for the negative
 parity states by OJH \cite{oka}. Now in this approach,
the vacuum expectation value of the  two interpolating fields
in Eq.~(\ref{c_soft}), $J_{N^*}$ being our new current with the 
covariant derivative, takes the form 
\begin{equation} 
\langle 0 | T J_{N^*} (x) {\bar J}_N (0) | 0 
\rangle \sim A \not \!z \not\!q + B \not \!z\ .
\end{equation}
Therefore, the structure corresponding to ${\gamma_5 \not \!z \not\!q}$ 
does not vanish in the soft-pion limit, and there is no contamination coming 
from the nucleon double pole term proportional to $g_{\pi {\rm NN}}$.  
As one can
see below, the phenomenological side also contains this structure.  
Therefore, since we have the well-defined structure on both sides,
it now makes sense
to compare the two sides in order to extract the physical parameter, 
$g_{\rm \pi {\rm NS}_{11}}$.  

By using the effective Lagrangian for $\pi$-N-S$_{11}$ interactions 
\begin{equation}
{\cal L} = g_{\rm \pi {\rm NS}_{11}} {\bar {\rm S}}_{11} \tau \cdot \pi N\ 
+ {\rm h. c.}\ ,
\end{equation}
one obtains the phenomenological side of the correlation function, 
Eq.~(\ref{corr}),
\begin{eqnarray}
\Pi^\pi_{\rm phen} = -i \gamma_5 \not \!z \lambda_{N^*} \lambda_{N} 
g_{\rm \pi {\rm NS}_{11}} \left [ { q^2 + M_N M_{N^*} 
\over (q^2 - M_N^2) (q^2 -M_{N^*}^2)} + \not\!q {M_N + M_{N^*} 
\over (q^2 - M_N^2) (q^2 -M_{N^*}^2)} \right ]\ .
\end{eqnarray}
Note that the dimension of the coupling $\lambda_{N^*}$ is one order higher
than $\lambda_{N}$ due to the covariant derivative introduced
in our S$_{11}$ current.  Because of this difference, only odd-dimensional
operators contribute to the OPE side in contrast to Ref.~\cite{oka} where
even-dimensional operators contribute to their sum rule.
The term proportional to ${i \gamma_5 \not\!q \not \!z} $ is
\begin{eqnarray}
\lambda_{N^*} \lambda_{N} 
{ g_{\rm \pi {\rm NS}_{11}} \over M_{N^*} - M_{N}} \left[ {1 \over 
q^2 -M^2_{N^*}} - {1 \over q^2 -M^2_{N}} \right ]\ .
\end{eqnarray}
The spectral density $\rho (s)$ in the spectral representation is basically
the imaginary part
of this function.  So, after the Borel transformation
$ \left [\Pi (M^2) =\int ds e^{-s/M^2}
\rho (s) \right ]$,  the correlator for this specific
structure becomes 
\begin{eqnarray}
{\hat \Pi}^{\rm phen} (M^2) = \lambda_{N^*} \lambda_{N} 
{g_{\rm \pi {\rm NS}_{11}}
\over  M_{N^*} - M_{N}} \left [ e^{-M_{N}^2/M^2} - e^{-M_{N^*}^2/M^2} 
\right ]\ .
\label{piphen}
\end{eqnarray}
Here $M$ denotes the Borel mass.  Since the phenomenological side 
at this structure is proportional to 
the difference between the positive and negative-parity states,
the continuum contribution to the sum rule should be small.  
However, we will also present the results with the continuum below.

The theoretical side of the correlation function, Eq.~(\ref{corr}),  is 
obtained by calculating the
two-point function in the operator product expansion (OPE). [For 
technical or general review, Ref.~\cite{qsr} will be useful.] 
To proceed,
the following replacements in the lowest order of the short-distance expansion
are useful,
\begin{eqnarray}
&&\langle 0 |q^\alpha_a (x)\ {\bar q}^\beta_{b} (0)|\pi^0 \rangle  
                                    \rightarrow i \gamma_5 
                                    \delta_{ab} \delta_{\alpha \beta} 
                                    {1 \over 12} 
                                   \langle 0 | {\bar q} i \gamma_5 q |\pi^0 
                                   \rangle\ , \nonumber \\
&&\langle 0 |q^\alpha_a (x)\ g_s G^A_{\mu\nu} (0) {\bar q}^\beta_{b} (0)
|\pi^0 \rangle
\rightarrow
                        (i\gamma_5 \sigma_{\mu\nu})^{\alpha \beta} t^A_{ab}
                        {1 \over 192} 
                        \langle 0 | 
                        {\bar q} i \gamma_5 g_s \sigma \cdot {\cal G} q
                        |\pi^0 \rangle\
\end{eqnarray}
where the gluon field tensor is defined as 
${\cal G}_{\mu \nu} \equiv  G^A_{\mu \nu} t^A$ and the 3$\times$3 matrix
$t^A$ is related to Gell-Mann matrix via $t^A= {\lambda^A \over 2}$.

The OPE expression for the correlator contributing to the specific
structure of our concern
is readily calculated up to dimension 7 as
\begin{eqnarray}
&& -(1+t) { \langle 0 | {\bar u} i \gamma_5 u |\pi^0 \rangle \over 48 \pi^2} 
q^2 
{\rm ln} (-q^2)\  \nonumber \\
&&+{{\rm ln}(-q^2) \over 96 \pi^2} \left [ (1+t) 
\langle 0 | {\bar u} i \gamma_5 g_s \sigma \cdot {\cal G} u |\pi^0 \rangle\
-(1-t)
\langle 0 | {\bar d} i \gamma_5 g_s \sigma \cdot {\cal G} d |\pi^0 \rangle\
\right ]\  \nonumber \\
 && +
{1+t \over 288} \left \langle {\alpha_s \over \pi} {\cal G}^2 \right \rangle 
\langle 0| {\bar u} i \gamma_5 u |\pi^0\rangle\ .
\end{eqnarray}
Note that there are
three possible ways to construct the dimension 5 operator in the fixed point
 gauge; 
(1) by taking the gluon field strength  ${\cal G}_{\mu \nu}$ 
from the quark-propagator 
(2) by taking ${\cal G}_{\mu \nu}$ from $z \cdot D$
(3) by taking the term 
containing two covariant derivatives in the short distant expansion of
$\langle 0 |q^\alpha_a (x)\ {\bar q}^\beta_{b} (0) |\pi \rangle$. 
It turns out that u-quark contribution from the second possibility 
cancels with the
one from the third possibility.  Also the nonzero contribution
from d-quark comes only from the second possibility.  In Ref.[x], the
OPE expression is symmetric under u-d quark exchange.  In our case,
that symmetry does not appear anymore
because of the covariant derivative introduced on the d-quark in 
Eq.~(\ref{nstar}).  

We then use the soft-pion theorem to write the vacuum to pion condensates in 
terms of the vacuum to vacuum condensates.  Then after the Borel 
transformation,  we obtain the OPE expression for the 
term proportional to ${i \gamma_5 \not\!q \not \!z} $,
\begin{eqnarray}
{\hat \Pi^{\rm ope}}(M^2) = {\langle {\bar q} q \rangle \over 16 \pi^2 
f_\pi } \left [ {M^4 \over 3} (1+t) + {2 \over 3} \lambda^2_q M^2
-{\pi^2 \over 18} (1+t) \left \langle {\alpha_s \over \pi} {\cal G}^2 
\right \rangle \right ]\ .
\label{piope}
\end{eqnarray} 
Here $\langle {\bar q} q \rangle = 
\langle {\bar u} u \rangle = 
\langle {\bar d} d \rangle $, and
$\lambda^2_q$ is the parameter associated with the quark-gluon condensate via
$\langle {\bar q} g_s \sigma \cdot {\cal G} q \rangle \equiv 
2 \lambda^2_q \langle {\bar q} q \rangle$.  The pion decay constant
$f_\pi = 0.093 $ GeV.  
The coupling $g_{\pi {\rm NS}_{11}}$ is obtained by equating this 
with Eq.~(\ref{piphen}). 

Now we discuss the case for the $\eta$-N-S$_{11}$ coupling.  In this case,
we consider the correlation function similar to Eq.~(\ref{corr}) with
$\pi^0$ replaced by $\eta$.  In the following,  
we will assume that there is no mixing from the singlet $\eta'$.  From the 
effective Lagrangian  
\begin{equation}
{\cal L} = g_{\rm \eta {\rm NS}_{11}} {\bar {\rm S}}_{11} \tau^0  \eta N\ 
+ {\rm h. c.}\ ,
\end{equation}
the phenomenological side is obtained from Eq.~(\ref{piphen}) by 
simply replacing 
$g_{\pi {\rm NS}_{11}}$ with $g_{\eta {\rm NS}_{11}}$.
Also
the OPE side in the soft $\eta$ limit is given similarly as in
  Eq.~(\ref{piope}).  The difference
is that the condensate involving d-quark 
now has the same sign as the 
condensates involving u-quark.  This follows after applying the soft-meson
theorem to write the vacuum to $\eta$ condensates in terms
of the vacuum to vacuum condensates,  
So the coupling
$g_{\eta {\rm NS}_{11}}$ in the QCD sum rule is given by 
\begin{eqnarray}
g_{\eta {\rm NS}_{11}} &=& {M_{N^*} -M_N \over \lambda_{N^*} \lambda_N}
\left [ e^{-M_N^2/ M^2} - e^{-M_{N^*}^2/M^2} \right ]^{-1}\nonumber \\
&\times& {\langle {\bar q} q \rangle \over 16 \pi^2
\sqrt {3} f_\pi } \left [ {M^4 \over 3} (1+t) + {2 t \over 3} \lambda^2_q M^2
-{\pi^2 \over 18} (1+t) \left \langle {\alpha_s \over \pi} {\cal G}^2
\right \rangle \right ]\ .
\label{etasum}
\end{eqnarray}
Compared with the $g_{\pi {\rm NS}_{11}}$ sum rule, the term containing 
$\lambda^2_q$ has the parameter $t$ whose value is determined from
the S$_{11}$ sum rule using the finite energy sum rule (FESR)~\cite{hung}.   

The couplings, $\lambda_{N^*}$ and $\lambda_{N}$, are determined from the
nucleon and S$_{11}$ sum rule respectively.  However, since the sign of 
the couplings  are not known from these sum rules,
we do not know 
the sign of $g_{\eta {\rm NS}_{11}}$ or $g_{\pi {\rm NS}_{11}}$ in
this approach.
The nucleon mass and the coupling of $J_{N}$ to the physical nucleon are 
\begin{equation}
|\lambda_{N}| = 0.0258~~{\rm GeV}^3 \;; \quad M_{N} = 0.977~~{\rm GeV} \ 
\end{equation}
which are obtained from the conventional nucleon sum rule~\cite{qsr,griegel} 
using the QCD parameters
\begin{eqnarray}
\langle {\bar q} q \rangle = -(0.23\ {\rm GeV})^3\;; \quad
\left \langle {\alpha_s \over \pi} {\cal G}^2 \right \rangle = 
(0.35\ {\rm GeV})^4\ .
\label{par}
\end{eqnarray}
  One of the results found in Ref.~\cite{hung} is that the mass splitting 
between nucleon and S$_{11}$ is due to the appearance of the quark-gluon condensate  
[$\langle {\bar q} g_s \sigma \cdot {\cal G} q \rangle \equiv 
2 \lambda^2_q \langle {\bar q} q \rangle$] whose value however is not well
known.  
Currently, the associated parameter is known within the range of
$0.4~{\rm GeV}^2~\le \lambda_q^2~ \le 1~{\rm GeV}^2$.  Therefore, for a 
given value of $\lambda_q^2$ within this range, we first determined the 
parameters $t$, $\lambda_{N^*}$ and $M_{N^*}$ using the 
S$_{11}$ sum rule~\cite{hung} and listed in Table~\ref{tab}.
  For each parameter set, we plot the
Borel curves for the couplings[Eq.~(\ref{etasum}) for $g_{\eta {\rm NS}_{11}}$
 and the corresponding formula for $g_{\pi {\rm NS}_{11}}$] and 
take the minima of the curves to obtain the couplings.  These values are also 
listed in the table.

The results for $g_{\eta {\rm NS}_{11}}$ are close to its empirical value of 2.
As the parameter $\lambda_q^2$ increases, $g_{\eta {\rm NS}_{11}}$  
increases slightly.  The continuum, if included in our formalism [see for example 
Ref.~\cite{hung,griegel}.],
increases the value by about 30 \%  as indicated by the numbers in the
parenthesis.  Note that, in Eq.(\ref{etasum}), the
first two terms of the OPE have the same sign in all cases we consider here. 
Therefore, the stability of the Borel curve is driven by these two terms 
and the relative contributions from higher dimensional operators are
small, providing only the range of minimum Borel mass.  Indeed, the dimension 7 operator contributes to the sum rule less
than 3 \% in the vicinity of the minimum of the Borel curve in all the
 cases we considered.     
Thus, the quark-gluon condensate is the crucial part of the
sum rule.

On the other hand, our results for $g_{\pi {\rm NS}_{11}}$ are
not as  good as the
$g_{\eta {\rm NS}_{11}}$ case.  
The value of  $g_{\pi {\rm NS}_{11}}$ is found to be closed to its empirical 
value of 0.7 when $\lambda_q^2 =0.5$ GeV$^2$.   Also at this value of
$\lambda_q^2$, the
coupling strength of the interpolating field to the physical S$_{11}$ is
the greatest. As the value of  $\lambda_q^2$ decreases from 0.5 GeV$^2$,
our prediction slightly increases.
However, for  $\lambda_q^2$ greater than 0.5 GeV$^2$, there is no stable 
point of the Borel curve
from which we can extract the value of $g_{\pi {\rm NS}_{11}}$. 
This can be explained by looking at the OPE expression for 
$g_{\pi {\rm NS}_{11}}$ given in Eq.~(\ref{piope}).  For large $\lambda_q^2$s,
the contribution from the quark-gluon condensate
has the opposite sign of the one from the quark condensate. Thus,
the two dominant terms in the OPE side cancel each other so that
the contribution from the higher dimensions is not negligible.  Indeed, for
$\lambda_q^2$ greater than 0.5 GeV$^2$, the sum of the first two terms 
is comparable
in magnitude with the dimension 7 term except for the region of high Borel
mass, at which the uncertainty coming from the continuum overwhelms.  
To get a stable Borel curve in this case, one should calculate the 
contribution from even higher dimensional operators.

The importance of the quark-gluon condensate in our prediction is also 
physically  clear.  Comparing to the usual nucleon sum rule,
the additional piece of the S$_{11}$ sum rule~\cite{hung} is the quark-gluon
condensate. Therefore, it is natural to expect that this condensate participates
in the sum rule calculation of coupling constants containing the 
S$_{11}$ resonance. Indeed, in our approach, 
the difference between $g_{\pi {\rm NS}_{11}}$ and $g_{\eta {\rm NS}_{11}}$
originates from this condensate.  Of course, more reliable prediction
can be made if the possible value for the quark-gluon condensate is narrowed
down further.   It should also be important to go beyond the soft-meson limit,
 which should especially be important for the $\eta$.  Also, the effect
coming from the mixing to $\eta'$ should be investigated.

In summary, we have determined the couplings, $g_{\pi {\rm NS}_{11}}$ and
$g_{\eta {\rm NS}_{11}}$ within the conventional QCD sum rule approach using
the interpolating field containing the covariant derivative.  
Due to the specific structure of our interpolating field, only odd-dimensional
operators contribute to our sum rule.  
Our prediction for
$g_{\eta {\rm NS}_{11}}$ agrees well with its empirical value and
relatively insensitive to the barely known QCD parameter $\lambda_q^2$.
 For the $g_{\pi {\rm NS}_{11}}$ case, however, the prediction  depends upon
the precise value of the quark-gluon condensate.
We discussed the role of the quark-gluon operator in our prediction.

\acknowledgments
SHL was supported in part by the by the Basic Science Research
Institute program of the Korean Ministry of Education through grant no.
BSRI-96-2425, by KOSEF through the CTP at Seoul National University.  The work
of SHL and HK was also supported by KOSEF, grant no. 971-0204-017-2.

\begin{table}
\caption{Our prediction for $g_{\pi {\rm NS}_{11}}$ and 
$g_{\eta {\rm NS}_{11}}$ at given $\lambda^2_q$.  The numbers in
parenthesis are when the continuum contributions are included.
The optimal $t$ is obtained from finite energy sum rule and
$\lambda_{N^*}$ and $M_{N^*}$ are obtained from the most
stable Borel plateau from Ref.~\protect{\cite{hung}}.}

\begin{center}
\begin{tabular}{cccccc}
 $\lambda^2_q$ (GeV$^2$) & $t$ from FESR &$|\lambda_{N^*}| 
({\rm GeV}^4)$&$M_{N^*}$ (GeV)
&$|g_{\pi {\rm NS}_{11}}|$&$|g_{\eta {\rm NS}_{11}}|$  \\
\hline\hline
0.435 & 7.94 & 0.044 & 1.535 & 1.55 (2.11) & 1.99 (2.67) \\
0.48 & 41.39 & 0.22  & 1.58 & 1.28 (1.77) & 2.21 (2.93) \\
0.5 & -47.52 & 0.24 & 1.59 & 1.1 (1.51) & 2.27 (3.03) \\
0.6 & -4.59 & 0.023 & 1.59 & N/A &  2.43 (3.25) \\
0.7 & -2.91 & 0.015 & 1.53 & N/A & 2.44 (3.27) \\
0.8 & -2.40 & 0.013  & 1.48 & N/A & 2.47 (3.3) \\
\end{tabular}
\end{center}
\label{tab}

\end{table}

\end{document}